# Realization and preliminary measurements on a 94 GHz SIS mixer

L. Oberto[1], N. De Leo[1,2], M. Fretto[1], A. Tartari[3], L. Brunetti[1], V. Lacquaniti[1]

[1] *Istituto Nazionale di Ricerca Metrologica, strada delle Cacce 91, 10135, Torino, Italy,
{l.oberto,n.deleo,m.fretto,l.brunetti,v.lacquaniti}@inrim.it*
[2] *Politecnico di Torino, corso Duca degli Abruzzi 24, 10124, Torino, Italy*
[3] *Università di Milano "Bicocca", Piazza della Scienza 3, 20126, Milano, Italy, A.Tartari@mib.infn.it*

*Abstract*-In this paper we present the realization and a preliminary characterization of a SIS based receiver. It has been developed for the MASTER experiment that consists in a three-band SIS receiver (94, 225 and 345 GHz) for astrophysical observations through the atmospheric windows available at high altitude dry sites. The measurements performed establish an upper limit to the overall receiver noise temperature. A comparison has been tried with the MASTER requirements and with state of the art results. A noise figure of 110 K has been obtained at 94 GHz, about 22 times the quantum limit.

**Keywords**: low-noise detection, SIS-mixer, milimeter-wavelength, radio-astromony.

## 1. Introduction

A fundamental task of cosmology is the understanding of the "Big Bang". Its residual, the cosmic background radiation (CMB) at 2.73 K discovered by Penzias and Wilson in 1963 [1], can conveniently be studied around 100 GHz. Important results have been recently obtained by the BOOMERANG [2] and the WMAP [3] collaborations using traditional techniques. More sensitive and less noisy instruments may allow obtaining better measurements.

Another central question in the study of the early Universe is understanding when and how galaxies are formed. Studies of stars in nearby galaxies indicate that first stars were formed in a dust-free environment but, since the birth of stars involves rapid production of heavy elements and dust, the radiation in the optical regime is obscured. This makes difficult the observation of young galaxies in the visible domain. Nevertheless, they are very bright at millimeter and sub-millimeter wavelengths (up to THz frequencies) where the radiation absorbed by the dust is re-emitted. A large amount of warm gas composed by molecules is also expected to be present and its emission lies in the sub-millimeter range. For these reasons, observations at millimeter and sub-millimeter wavelengths are essential to understand the origin of the universe.

Radioastronomical observations require extreme sensitivity but nature imposes an upper limit to the mankind ability in acquiring information. This limit is defined by the Heisenberg's indetermination principle. When the quantum mechanical aspect of nature appears, it limits the accuracy of measurements. Therefore, heterodyne detection using mixer based on Josephson Superconductor-Insulator-Superconductor (SIS) junctions as nonlinear element is widely diffused. In fact SIS mixers are able to approach the quantum noise limit that is, at 94 GHz, about 5K.

The development of SIS mixers requires very high technological standards. With this work we show a preliminary characterization of a SIS mixer designed and realized at the Istituto Nazionale di Ricerca Metrologica (INRIM, Torino, Italy) for the 94 GHz channel of the MASTER receiver.

### 1.1. The MASTER experiment

The MASTER experiment [4] is a collaboration between Università di Milano "Bicocca", Istituto Nazionale di Ricerca Metrologica (INRIM, formerly Istituto Elettrotecnico Nazionale, IEN "G. Ferraris"), Istituto Nazionale di Astrofisica (INAF) – Osservatorio Astrofisico di Arcetri, Consiglio Nazionale delle Ricerche (CNR) and Università di Roma "La Sapienza". It consists of a system of three heterodyne receivers for astrophysical observations at 94, 225 and 345 GHz through the atmospheric windows available at dry, high altitude sites (Italian Alps, Antarctic Plateau). The core of the system are SIS mixers realized with Nb/Al-AlO$_x$/Nb tunnel junctions.

The 94 GHz mixer is realized at INRIM while the 225 GHz unit is manufactured at the Department of Astronomy of the University of Massachussets Amherst (USA) and the 345 GHz receiver is made by the KOSMA laboratory at the Physics Institute of the University of Cologne, Germany.

The three SIS units are cooled below the Niobium critical temperature $T_C$ = 9.25 K using a liquid Helium bath contained into a standard Nitrogen-Helium Dewar. A solution based on a Cryocooler-liquid Helium hybrid system is also feasible.

The three Local Oscillator signals are produced by two Gunn Oscillators: the first tunable between 88 and 100 GHz and the second one between 73 and 87.5 GHz. The former is used for the 94 GHz receiver while the latter, splitted in two parts multiplied, respectively, by 3 and 4, is used for the 225 and the 345 GHz receivers. A Phase Lock system stabilizes the LO signal frequency to allow spectroscopic measurements. In Table 1 the MASTER requirements are listed.

|  | 94 GHz | 225 GHz | 345 GHz |
|---|---|---|---|
| IF frequency / GHz | 1.5 | 1.5 | 1.5 |
| Inst. Bandwidth / GHz | 0.6 | 0.6 | 0.6 |
| DSB System noise Temp / K | 100÷130 | 120÷150 | 140÷170 |
| Tuning bandwidth / GHz | 10 | 20 | 35 |

Table 1. MASTER requirements.

## 2. SIS mixer fabrication

In our mixer we have used Nb/Al-AlOx/Nb Josephson junctions embedded in a niobium microstripline (see Figure 1). An optimal performance of the mixer device is achieved optimizing the quality of both SIS junction and the microstripline film which the junction is embedded in.

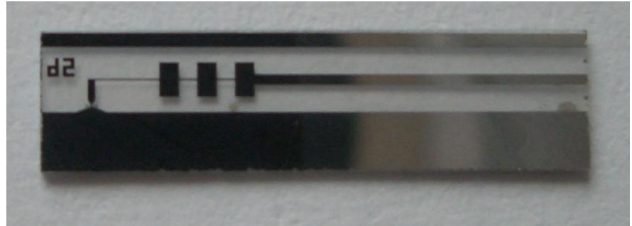

Figure 1. Picture of the MASTER 94 GHz SIS mixer chip.

It is well-established that the highest mixing frequency depends on parameters such as the junction energy gap and width but also the strip line cut-off frequency. Moreover the mixing properties of the SIS junction can be enhanced provided that the current-voltage characteristic presents almost ideal non-linearity at the working temperature (4.2 K).

The SIS trilayer has been deposited on a quartz substrate 0.3 mm thick. The deposition of the Nb base and top electrode layers (120 nm thick) has been carried out at a rate of 0.65 nm/s. After a waiting time of 30 min, to allow Nb base layer to cool and relax its internal stress, the Al film (5 nm thick) has been deposited at 0.1 nm/s and immediately oxidized in pure oxygen. The deposition rate and the thickness of this film are quite specific of these junctions. In fact, while a highly non-linear hysteretic I-V characteristic was obtained with the mentioned values, the properties of these junctions can be turned in a controllable way into fully non-hysteretic, if the aluminium thickness is raised to 30 nm or more, and the deposition rate is made > 1nm/s [5]. This is due to the fundamental role of the surface roughness of this layer and its dependence on thickness and deposition rate. The oxidation parameters were chosen and controlled to reach the desired value of critical current densities. Therefore, different values of oxidation exposure were used depending on the mixing frequency: in particular for a 94 GHz mixer, 10 000 − 12 000 Pa · s to obtain a current density of 1-1.5 kA/cm$^2$.

The deposition of this SIS structure, previously made by a RF High Vacuum sputtering system, is now carried out in a DC Ultra-HV sputtering machine with two different deposition chambers for Nb and Al. The base vacuum for the Nb, the most influencing film, is in the range of $10^{-8}$-$10^{-9}$ Pa. Junction area were 5 x 5 μm$^2$, corresponding to a capacitance of about 2 pF.

The high structural quality of the Nb films, showing monocrystal growth with minimal misorientation of the grains, has been confirmed comparing electrical measurements of these junctions with analogous junctions fabricated by RF HV sputtering. In particular, measurement of sub-nanocurrent at T < 4.2 K in the sub-gap region has demonstrated the absence of any defects (such as pinholes) in the nanometric oxide barrier, grown on the Nb/Al bilayer obtained by DC UHV sputtering.

Concerning the microstripline, constituted by the niobium base layer, its electrical properties can be further improved increasing the thickness of the film to 200nm. The last important aspect is the electrical and mechanical contact of the SIS mixer to the microwave mounting. This connection has been recently optimized improving the quality of the Ti/Au bilayer (30/300 nm) used to make electrical contacts, obtained by electron gun evaporation in High Vacuum.

### 3. The MASTER 94 GHz receiver

Figure 2 shows a schematic of the MASTER 94 GHz SIS receiver. It consists of a cryogenic section containing the SIS mixer with a built-in IF filter, a cryogenic isolator and a first IF HEMT amplifier (35 dB). A warm section includes the IF electronics with two other amplifiers (40 and 20 dB), the bias (not shown in the picture) and the Data Acquisition systems. The Local Oscillator is coupled to the mixer using a feed horn and a beam splitter through whom the astronomical signal is also conveyed to the Josephson junction. A variable attenuator can also be included in the warm IF electronics to prevent amplifier saturation.

The feed horn output port is a WR10 waveguide whereas the mixer mount consists in a tapered waveguide input terminated with a movable back-short for better coupling [6]. The IF output of the mixer is fitted with a coaxial SMA connector.

The IF electronics is manufactured by INAF - Osservatorio Astrofisico di Arcetri, the DAQ and bias electronic is made by Università di Milano "Bicocca" while the SIS mixer is realized by INRIM.

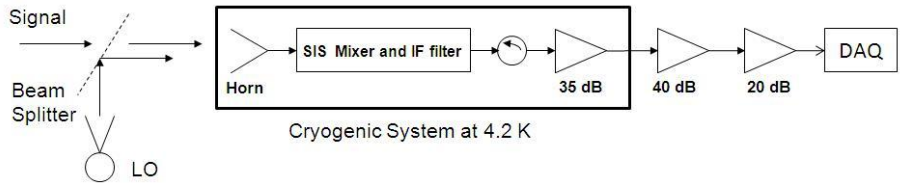

Figure 2. Schematization of the 94 GHz MASTER receiver.

### 4. Measurement results

The DC I-V characteristic of the produced devices has been recorded to verify the junctions quality. In Figure 3 typical un-pumped and pumped curves can be seen. The pumped one is obtained irradiating the junction with the LO signal and it shows the first photon step [7].

The mixer mount has been manufactured after simulation studies that predicted a reflection coefficient, at the input port, of about 0.4 in a 2 GHz-wide frequency band centred around the tuning frequency of 94 GHz [6]. To verify this, we have performed scattering parameter measurements with a Vector Network Analyzer at room temperature. Results are shown in Figure 4 for different back-short distances from the SIS juncions: it can be seen that the required figures are obtained around 91 GHz. This is not surprising because of mechanical inaccuracies and because, at room temperature, Niobium behaves as a resistor so that the electromagnetic characteristic of the structure slightly changes. Nevertheless, we expect to obtain optimum results around 94 GHz at cryogenic temperatures as predicted.

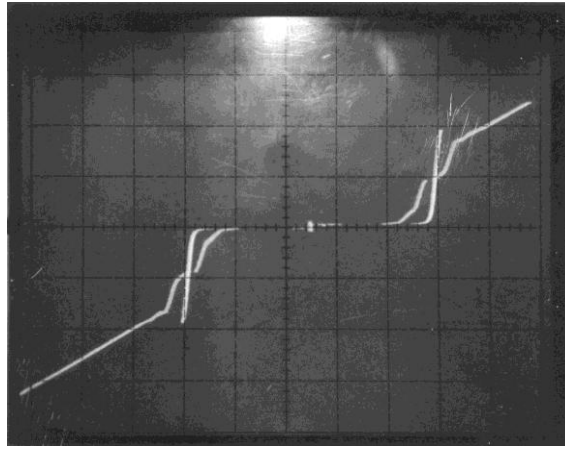

Figure 3. Mixer I-V pumped and un-pumped curves. The horizontal scale is 1 mV/div; the vertical one is 0.1 mA/div.

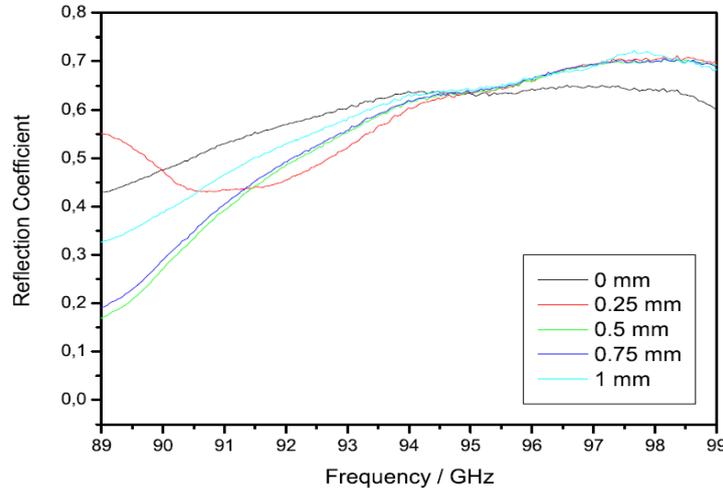

Figure 4. Reflection coefficient measurement at the input port of the mixer mount for different back-short position. A colour version of the graph can be found in the on-line version of the paper.

Noise temperature measurements have been, then, performed using the *Y-factor* technique [8]. It allows the determination of the overall noise temperature $T_N$ of the system, if we let it receive the wide band emission of two blackbody sources of known temperature $T_H$ and $T_L$, with the following equation:

$$T_N = \frac{T_H - YT_L}{Y - 1}, \qquad (1)$$

where

$$Y = \frac{P_{out,H}}{P_{out,L}} \qquad (2)$$

in which $P_{out, H}$ and $P_{out, L}$ are the receiver output powers when it detects the signal from the high and low temperature blackbodies, respectively. For these measurements we have chosen the ambient and the liquid Nitrogen temperature. In Figure 5 the results are presented. As expected, optimum performances are in the 2 GHz band around 94 GHz and the best figure is 110 K, compliant to the MASTER requirements of Table 1. No attempt has been made in order to evaluate the uncertainties of these measurements since the device is a prototype with the only aim to demonstrate the INRIM ability to realize superconductive Josephson mixers. The receiver electronics is also not fully characterized, yet. However, since a better coupling between the mixer chip and the incoming radiation can be achieved, we claim that the real receiver noise will be certainly lower, so that 110 K is a realistic upper bound.

The quantum noise limit at 94 GHz is $hf/k_B \cong 5$ K ($h$ being the Plank constant, $f$ the frequency and $k_B$ the Boltzmann constant) while state of the art data reports noise valuess between 4 and 10 times $hf/k_B$ [9]. Therefore our result $T_N < 22\ hf/k_B$ is promising, even if demonstrating cutting edge performances was not our target.

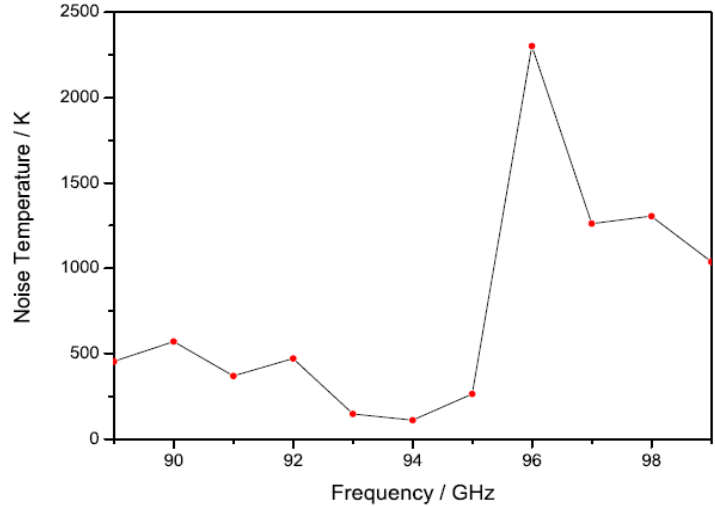

Figure 5. Preliminary results of the noise temperature of the MASTER 94 GHz prototype receiver. A colour version of the graph can be found in the on-line version of the paper.

The typical behaviour of the recorded signals is presented in Figure 6. The black line is the DC I-V characteristic, the red and the blue lines represent the hot and cold load downconverted power peaks. In green the dynamic resistance trend is shown; it is similar to the ones of the conversion peaks as foreseen by the Tucker's Quantum Mixing Theory [7]. All curves are represented versus the junction bias voltage.

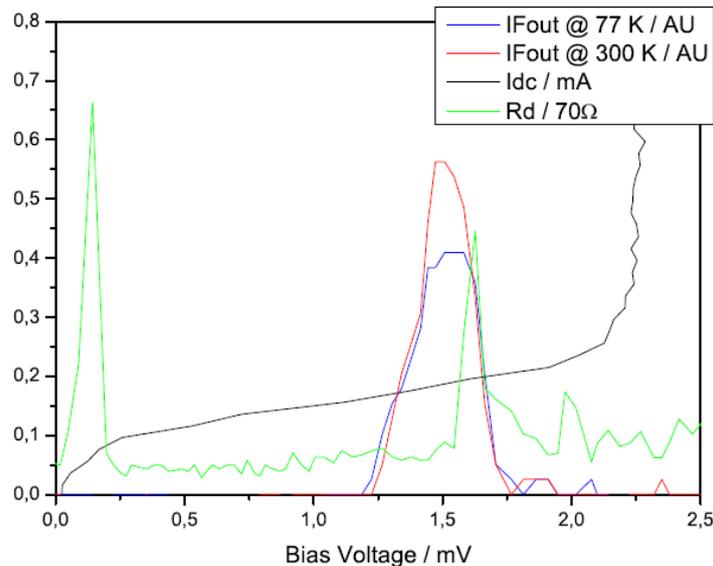

Figure 6. IF output hot and cold load down-converted signals, junction DC I-V characteristic and dynamic resistance (scaled by a factor of 70 to fit the graph scale). A colour version of the graph can be found in the on-line version of the paper.

## 5. Conclusions

In this work we have presented preliminary results concerning the characterization of a prototype for the MASTER experiment 94 GHz receiver based on a superconductive SIS mixer developed at INRIM. Measurements of its noise temperatures have been performed with promising results. They can be regarded as upper bound for the final overall receiver noise temperature that is compliant with the experiment requirements. Moreover, this limit is not far from the actual state of the art in the field.

The realization of the device is also described with attention to the optimization of crucial features along with a DC characterization of its Josephson junctions.

Finally, we have demonstrated the INRIM ability to design and realize SIS mixers usable for practical astrophysical observations.

## Acknowledgements

Authors want to thank for their valuable contribution Dr. Franco Delpiano and Dr. Eugenio Monticone. They have given a considerable help in the optimization of the soldering procedure and deposition of the Ti/Au bilayer. Our gratitude goes also to Dr. Adriana Manenti for paper revision.